\pdfoutput=1 

\documentclass[sigplan,screen,10pt]{acmart}

\setcopyright{acmlicensed}
\copyrightyear{2024}
\acmYear{2024}
\acmDOI{XXXXXXX.XXXXXXX}

\acmConference[Onward'24]{2024 ACM SIGPLAN International Symposium on New Ideas, New Paradigms, and Reflections on Programming and Software}{October 23--25,
  2024}{Pasadena, California, United States}
\acmISBN{978-1-4503-XXXX-X/18/06}

\usepackage[utf8]{inputenc}
\usepackage{booktabs}
\usepackage{hyperref}
\usepackage{listings}
\usepackage[normalem]{ulem}

\lstdefinestyle{statute}
{
basicstyle=\small \rmfamily \linespread{0.25}\selectfont, columns=flexible,
breaklines=true, breakatwhitespace=true,  breakindent=0pt, extendedchars=true, 
literate={§}{{\S}}1
{—}{{---}}1
{“}{{``}}1
{”}{{''}}1,
escapechar=@
}

\lstdefinestyle{llm}
{
basicstyle=\footnotesize \ttfamily \selectfont, columns=flexible,
breaklines=true, breakatwhitespace=true, breakindent=0pt, extendedchars=true, 
literate={§}{{\S}}1
{—}{{---}}1
{“}{{``}}1
{”}{{''}}1,
escapechar=@
}

\usepackage{amsthm}
\newtheorem{example}{Example}

\newcommand{\code}[1]{\texttt{#1}}

 \definecolor{boxcolor}{RGB}{238, 223, 204} %

\title{Software Engineering Methods For AI-Driven Deductive Legal Reasoning}

\author{Rohan Padhye}
\affiliation{%
  \institution{Carnegie Mellon University}
  \city{Pittsburgh, PA}
  \country{United States}}
\email{rohanpadhye@cmu.edu}


\begin{abstract}
The recent proliferation of generative artificial intelligence (AI) technologies such as pre-trained large language models (LLMs) has opened up new frontiers in computational law. An exciting area of development is the use of AI to automate the deductive rule-based reasoning inherent in statutory and contract law. This paper argues that such automated deductive legal reasoning can now be viewed from the lens of software engineering, treating LLMs as interpreters of natural-language programs with natural-language inputs. We show how it is possible to apply principled software engineering techniques to enhance AI-driven legal reasoning of complex statutes and to unlock new applications in automated meta-reasoning such as mutation-guided example generation and metamorphic property-based testing.
\end{abstract}

\begin{document}

\maketitle

\section{Introduction}

Numeric computations in law, regulation, and contracts govern many aspects of our personal and professional lives---owing taxes, qualifying for student financial aid, receiving social security benefits, breaking a rental lease, paying down a credit card balance, demanding compensation for a delayed flight, and bequeathing property in a will, to name a few examples. In each of these cases, the logic of such computation is encoded in legally-binding natural language documents of sometimes monstrous complexity. The process of establishing how such legal rules apply to specific situations usually falls into the category of \emph{deductive legal reasoning}~\cite{Ellsworth05-LegalReasoning} (as opposed to \emph{analogical} or case-based legal reasoning that draws upon precedents set by judges).

As such, deductive legal reasoning is highly mechanical when all the relevant facts are precisely known. Yet, this is by no means straightforward. For example, Appendix~\ref{app:irc} lists specific parts of the the \emph{Internal Revenue Code} (IRC), the body of statutory law that deals with U.S. federal taxation. Consider the following (well-specified) problem:

\begin{example}
\label{ex:interpretation}
Alice and Bob are married and filing taxes jointly for the year 2018. Alice was born on 1/1/1981 and Bob was born on 12/30/1975. In 2018, Alice and Bob's adjusted gross income was \$216,350. Alice and Bob do not itemize their deductions and do not qualify for any deductions other than the standard deduction. What is their taxable income for 2018?
\end{example}

\begin{proof}[Solution Approach]
Although a seemingly simple question, correctly answering it requires applying multiple computational rules from the IRC\footnote{While there are numerous paper worksheets and software packages produced by tax preparation services, those are not the law. Someone has to manually interpret the statutes in the IRC, which is the ground truth specification. This paper concerns the underlying legal analysis.} (ref. Appendix~\ref{app:irc}): \hyperref[irc:63b]{§63(b)} and \hyperref[irc:63c1]{§63(c)(1)} to identify the calculation of standard deduction and taxable income; \hyperref[irc:63c2Ai]{§63(c)(2)(A)(i)} and \hyperref[irc:63c2C]{§63(c)(2)(C)} to determine the formula for basic standard deduction applicable to joint returns; \hyperref[irc:63c7Aii]{§63(c)(7)(A)(ii)} which overrides the value \$3,000 in the preceding formula with the value \$12,000; as well as realizing that several other rules such as \hyperref[irc:63c3]{§63(c)(3)} and \hyperref[irc:63f1]{§63(f)(1)} do not apply because Alice and Bob are not 65 years old in 2018, and that inflation adjustments decreed in \hyperref[irc:63c4]{§63(c)(4)} and \hyperref[irc:63c7B]{§63(c)(7)(B)} do not apply in 2018. The final answer is \$216,350-(2$\times$\$12,000)=\$192,350.
\end{proof}

The reasoning process thus involves evaluating the composition of multiple rules, resolving internal cross-references, handling exceptions that may invalidate seemingly applicable rules, and performing arithmetic calculations; crucially, however, this process assumes that all the facts (as they pertain to the rule-based reasoning) are clearly known and so it does not deal with subjective interpretation of potentially ambiguous terms~\cite{tta-tradevsbusiness}. Similar legal reasoning is required in other previously mentioned domains: for example, determining to what extent an insurance policy might cover a particular incident, or correctly executing a will of a decedent given specific circumstances about their assets and beneficiaries.

The complexity increases even further when considering the objectives of those drafting the relevant legislation or contracts, who may be interested in a form of \emph{meta-reasoning} of edge cases. For example, government legislators might wonder if the tax law introduces any loopholes, whereas a testator might want to ensure that their will robustly expresses their intentions even in circumstances where some of their beneficiaries pass before themselves. The meta-reasoning problem is one of validating a legal text to satisfy some desirable properties across the universe of hypothetical scenarios.

Naturally, researchers have attempted to automate deductive legal reasoning to reduce the risk of costly errors; however, classical approaches have focused on formalizing the rules using special logics or domain-specific programming languages~\cite{Lawsky16-FormalizingIRC, Lawsky17-DefaultLogic, Holzenberger20-SARA, Merigoux21-Catala, Goodenough22-ComputableContracts}---while these approaches are sound, the formalization process itself appears to be too tedious for these techniques to achieve widespread adoption.

The recent proliferation of generative artificial intelligence (Gen AI) technologies such as pre-trained large language models (LLMs) have made it possible to directly automate deductive legal reasoning using the natural language documents~\cite{Goebel23-COLIEE23, Nay24-LLMsTaxAttorneys, Blair-Stanek23-GPT3Statutory}, with limited but progressively improving success rates. For example, prompting Anthropic's Claude 3 Opus~\cite{claude} with a preface (ref. Fig.~\ref{fig:system-prompt}) and the text from Example~\ref{ex:interpretation} is usually sufficient to solve the problem.\footnote{All LLM prompts and corresponding LLM responses related to examples listed in this paper are available at \url{https://github.com/cmu-pasta/onward24}.}

In this paper, we argue that AI-driven legal reasoning is essentially a form of (noisy) program execution, where both the implementation code and the program input is represented in natural language---the language models are thus \emph{interpreters}. As such, it should be possible to understand, debug,  optimize, analyze, maintain, and evolve these documents \emph{as if they were programs} using well-studied techniques from the field of \emph{software engineering}. With the help of a running example from tax law, we propose several applications of software engineering to legal reasoning:

\begin{itemize}
    \item Streamlined interpretation using \emph{subroutine inlining}.
    \item Example generation using \emph{coverage analysis} and \emph{mutation testing}.
    \item Edge-case analysis using \emph{property-based testing}.
\end{itemize}

Our hope is that this paper opens up new research directions at the cross-sections of computational law, artificial intelligence, and software engineering. We anticipate that such research will enable the development of enhanced search and analysis tools to assist lawmakers and citizens alike in reasoning about complex legal documents.

\begin{figure}
\begin{lstlisting}[style=llm]
You are an expert legal reasoning system, capable of reading legal statutes and applying them to given scenarios by identifying the correct subset of sections that apply to the particular situation.

In this conversation, we are only going to refer the following statutes. Do not use your knowledge of any other piece of legislation not provided in this conversation. If any undefined section is referenced in a question, please refuse to answer the question. The statutes referenced in this conversation are as follows:

```
<Excerpts from the IRC as provided verbatim in Appendix @\ref{app:irc}@>
```
\end{lstlisting}
\caption{LLM system prompt for all examples in this paper.}
\label{fig:system-prompt}
\end{figure}

\section{Interpreting Substitutions via Inlining}
\label{sec:substitution-inlining}

One of the challenges with using LLMs for automated deductive legal reasoning is the cross-application of other sections with substituted values. For example, consider the task of determining the standard deduction for taxable year 2025, which requires calculating cost-of-living adjustments using hypothetical inflation data provided in Appendix~\ref{app:cpi}.\footnote{At the time of writing (June 2024), the inflation data for calendar year 2024 is not yet fully known and consequently the standard deduction for 2025 has not yet been published by the IRS; so, we are confident that an LLM will not be able to use external knowledge to solve this problem.}

\begin{example}
\label{ex:substitution}
    Alice and Bob are married and filing taxes jointly for the year 2025. What is the basic standard deduction available to them?
\end{example}

Solving this task correctly requires several considerations:
\begin{enumerate}
    \item For joint returns, \hyperref[irc:63c2Ai]{§63(c)(2)(A)(i)} declares the basic standard deduction to be 200\% of the value in \hyperref[irc:63c2C]{§63(c)(2)(C)}, which is stated to be \$3,000.
    \item \hyperref[irc:63c4]{§63(c)(4)} applies ``adjustments for inflation'' to the value in \hyperref[irc:63c2C]{§63(c)(2)(C))}, but later \hyperref[irc:63c7Bi]{§63(c)(7)(B)(i)} overrides this substitution by declaring that ``Paragraph (4) shall not apply'' for 2025. \label{step:63c4-not-apply}
    \item For 2025, \hyperref[irc:63c7Aii]{§63(c)(7)(A)(ii)} replaces the value ``\$3,000'' in \hyperref[irc:63c2C]{§63(c)(2)(C))} with the value "\$12,000". \label{step:63c2C-12000}
    \item As per \hyperref[irc:63c7Bii]{§63(c)(7)(B)(ii)}, the value ``\$12,000'' in \hyperref[irc:63c2C]{§63(c)(2)(C))} is further increased using a cost-of-living adjustment as per \hyperref[irc:1f3]{§1(f)(3)}, but applying the latter rule itself with a substitution of ``2017'' for ``2016''.
    \item \hyperref[irc:1f3A]{§1(f)(3)(A)} provides a formula for computing a cost-of-living adjustment using the C-CPI-U value for the preceeding calendar year (i.e., 2024) and the CPI for calendar year 2016 multiplied by the ratio of the C-CPI-U for 2016 to the CPI for 2016 as in \hyperref[irc:1f3B]{§1(f)(3)(B)}. However, \hyperref[irc:1f3C]{§1(f)(3)(C)} comes into play, changing the calculation because we are using ``2017'' instead of ``2016'', as determined in the previous step. Now, it appears we must only use the C-CPI-U values of 2024 and 2017; no multiplication is needed.\label{step:1f3A-2017}
    \item Plugging in the values from Appendix~\ref{app:cpi}, the adjustment is calculated to be 24.82\%.
    \item Applying the adjustment and the rounding specified in \hyperref[irc:63c7Bii]{§63(c)(7)(B)(ii)}, the value ``\$12,000'' from \hyperref[irc:63c2C]{§63(c)(2)(C))} is therefore determined to change to ``\$14,950''.\label{step:63c2C-15000}
    \item Finally, stepping back to \hyperref[irc:63c2Ai]{§63(c)(2)(A)(i)}, the basic standard deduction for a joint return is therefore 200\% of this adjusted value. So, the result is \$29,900.
\end{enumerate}

\begin{figure}
\begin{lstlisting}[style=llm]
<... System prompt (ref. Figure @\ref{fig:system-prompt}@) ... >

<... Text of problem (e.g., Example @\ref{ex:substitution}@) ... >

Let's think step by step as to which rules apply, and stop if you come across an applicable rule R that requires substitution of some value or phrase X in place of Y in rule S. In this case, do not attempt to apply any other rules or compute the final answer. Instead, I only want to perform a single substitute. Please only return a modified version of the relevant statute S with the substitution performed. End your response with the following:

Modified rule: S
Modified by: R
Updated text for rule S: <new text replacing Y with X>
\end{lstlisting}
\caption{LLM Prompt incorporating subroutine inlining.}
\label{fig:prompt-inlining}
\end{figure}

In our experiments, modern LLMs like Claude~3 Opus or GPT-4 struggle to solve this task correctly even with few-shot in-context learning~\cite{Brown20-GPT3} and using \emph{chain-of-thought} reasoning~\cite{Kojima22-ZeroShotCoT}. There are simply too many \emph{subroutines}, and the LLM's reasoning inevitably falters by using a wrong value or applying it in the wrong place. The end result is almost always incorrect---see Appendix~\ref{app:claude-inflation_baseline} for a sample response where Claude produces a wrong answer.

Taking inspiration from \emph{inlining} in software optimization~\cite{Davidson92}, we devised a prompting strategy where we asked the LLM not to calculate the final answer, but just perform one step of the substitution and reproduce a modified version of the statute verbatim (ref. Figure~\ref{fig:prompt-inlining}). For example, after step~\ref{step:63c4-not-apply} above, the LLM will return the text in Appendix~\ref{app:irc} except with \hyperref[irc:63c4]{§63(c)(4)} completely removed. After applying step~\ref{step:63c2C-12000} above, the modified version of the statute will say the following in place of \hyperref[irc:63c2C]{§63(c)(2)(C)} (emphasis added):

\begin{lstlisting}[style=statute]
    (2) Basic standard deduction 

        ...

        (B) @\textbf{\$18,000}@ in the case of a head of household (as defined in section 2(b)), or 
        
        (C) @\textbf{\$12,000}@ in any other case. 
\end{lstlisting}

\noindent Continuing in this way, applying step~\ref{step:1f3A-2017} above changes \hyperref[irc:1f3A]{§1(f)(3)(A)} to say  (emphasis added):

\begin{lstlisting}[style=statute]
    (3) Cost-of-living adjustment 

        For purposes of this subsection—

        (A) In general 

            The cost-of-living adjustment for any calendar year is the percentage (if any) by which—

            (i) the C-CPI-U for the preceding calendar year, exceeds 

            (ii) @\textbf{the C-CPI-U for calendar year 2017}@.
\end{lstlisting}

\noindent and so on. Using a modified version of the statute with \emph{inlined} substitutions in each subsequent step allows the LLM to slow down its chain of thought. With this process, Claude~3 Opus is able to correctly solve Example~\ref{ex:substitution} after seven iterations.\footnote{For this exercise, we manually ran the ``iterations'' by prompting the LLM multiple times and performing the substitution of rule $S$ and cleanup of rule $R$ as suggested by the LLM in its response (ref. Fig.\ref{fig:prompt-inlining}). In general, we expect that this step can be automated with simple text-processing scripts.}

\section{Diverse Example Generation}
\label{sec:example-generation}

Consider the task of generating one-paragraph example scenarios (as in Example~\ref{ex:interpretation}) such that certain rules in the statute apply in specific ways. This could be useful for documentation or education---for instance, the IRS embeds hundreds of examples across its various publications and form instructions in order to demonstrate how federal tax law applies across various circumstances. Additionally, high quality examples can also aid in training language models to perform more accurate legal reasoning---researchers have found that augmenting training data with additional examples improves accuracy on some tasks in the COLIEE dataset~\cite{Yoshioka21-DataAugmentation}.

In software engineering, the task analogous to \emph{diverse example generation} is that of constructing a high-quality suite of test cases. Software tests are useful in many ways including validation (i.e., checking whether a program behaves correctly), specification (i.e., writing a test before implementing functionality, as in \emph{test-driven development}~\cite{Beck03}), and documentation (i.e., the test code demonstrates the behavior of a certain application programming interface (API)).

\subsection{Code Coverage} 
\label{sec:coverage}

In software testing, the quality of a test suite is usually measured using criteria that track the diversity of program behaviors captured across all tests. The most widely used example is that of \emph{code coverage}, which measures the fraction of source code (often counted at the level of \emph{lines} of code) exercised across all tests. Code coverage is often used as an objective function in automated test-input generation tools, such as \emph{coverage-guided grey-box fuzzing}~\cite{Bohme16}.

Using LLMs, we can use the same principles as in software testing to measure the coverage of available examples and to generate more examples by targeting specific coverage criteria. For example, asking Claude~3 Opus to solve Example~\ref{ex:interpretation} with a coverage trace (ref. Appendix~\ref{app:claude-coverage}) correctly produces a list of rules whose conditions were satisfied in this example: 
``\hyperref[irc:63b]{§63(b)}, \hyperref[irc:63c1]{§63(c)(1)}, \hyperref[irc:63c7]{§63(c)(7)}, \hyperref[irc:63c7Aii]{§63(c)(7)(A)(ii)}, \hyperref[irc:63c2A]{§63(c)(2)(A)}, \hyperref[irc:63c2C]{§63(c)(2)(C)}''.
Moreover, we can ask the LLM to synthesize examples such that a particular rule is exercised in its coverage. For instance, asking Claude for an example such that \hyperref[irc:63f1B]{§63(f)(1)(B)} applies but \hyperref[irc:63f1A]{§63(f)(1)(A)} does not, produces:

\begin{example}
Charlie and Diane are married and filing taxes jointly for the year 2018. Charlie was born on 3/15/1985 and Diane was born on 8/22/1953. In 2018, Charlie and Diane's adjusted gross income was \$185,000. Charlie and Diane do not itemize their deductions and do not qualify for any deductions other than the standard deduction.
\end{example}

\begin{proof}[Explanation (as per Claude)] ``In this scenario, Diane has attained age 65 before the close of the 2018 tax year, so \hyperref[irc:63f1B]{§63(f)(1)(B)} applies, entitling Charlie to an additional standard deduction amount for his spouse. However, Charlie himself has not attained age 65 before the close of the 2018 tax year, so \hyperref[irc:63f1A]{§63(f)(1)(A)} does not apply''.
\end{proof}

\subsection{Mutation Testing} 
\label{sec:mutation-testing}

In software testing, a criticism of the \emph{code coverage} measurement of test-suite quality is that it only cares about whether certain lines of source code were \emph{exercised}, but not the meaning of those lines themselves. High-coverage test suites do not necessarily find more faults than low-coverage ones~\cite{Inozemtseva14}. Instead, what we really want is for our tests to be able to \emph{fail} when a software bug is introduced. \emph{Mutation testing}~\cite{DeMillo78, Papadakis19} measures test-suite quality by artificially injecting ``bugs'' and determining if tests fail on the buggy mutated programs. The objective of a test writer is then to produce new tests that would distinguish the behavior of the original program from that of the mutation. This idea has also been used to automate test-input generation with the objective of maximizing mutation-detection ability~\cite{Fraser10, Vikram23}. 

We can apply the same principles as in software mutation testing to generate examples that not only exercise certain legal rules, but also distinguish them from other incorrect rules. For example, we can ask an LLM to assume that \hyperref[irc:63f1B]{§63(f)(1)(B)} is modified to remove the second clause, as follows:

\begin{lstlisting}[style=statute]
        (B) for the spouse of the taxpayer if the spouse has attained age 65 before the close of the taxable year @\sout{and an additional exemption is allowable to the taxpayer for such spouse under section 151(b)}@.  
\end{lstlisting}

\noindent Then, we ask the model to synthesize an example such that \hyperref[irc:63f1B]{§63(f)(1)(B)} would apply with the modified rule but it would not apply with the original rule. Claude~3 Opus responds to this challenge with the following example:

\begin{example}
    Ellie and Frank are married, but they are filing their taxes separately for the year 2018. Ellie was born on 5/10/1980 and Frank was born on 2/28/1952. In 2018, Ellie's adjusted gross income was \$75,000 and Frank's adjusted gross income was \$50,000. Neither Ellie nor Frank itemize their deductions and they do not qualify for any deductions other than the standard deduction.
\label{ex:mutation-testing}
\end{example}

\begin{proof}[Explanation (as per Claude)] ``The modified rule [...] would apply to Frank because he has attained age 65 before the close of the 2018 tax year. This means Frank would be entitled to an additional standard deduction amount for being aged 65 or older, even though he and Ellie are filing separately. However, under the original rule, \hyperref[irc:63f1B]{§63(f)(1)(B)} would not apply to Frank because it requires not only that the spouse has attained age 65, but also that ``an additional exemption is allowable to the taxpayer for such spouse under section \hyperref[irc:151b]{§151(b)}.'' [...] In this scenario, Frank does have gross income, so even though he is over 65, the original \hyperref[irc:63f1B]{§63(f)(1)(B)} would not apply to him because Ellie cannot claim an additional exemption for him under \hyperref[irc:151b]{§151(b)}."
\end{proof}

Mutation testing thus helps with generating examples that not only exercise specific parts of relevant statutes but also provide a basis for reasoning about why certain rules are written in a certain way. 

\section{Metamorphic Property-Based Testing}
\label{sec:pbt}

Property-based testing, originally introduced in \emph{Quickcheck} for Haskell~\cite{Claessen00}, is a technique for writing generic tests that express what properties a function should satisfy given \emph{any} imaginable input. Given a function $f : X \rightarrow Y$ and a predicate $p : X \times Y \rightarrow \mathrm{bool}$, we want to ensure that $\forall x \in X: p(x, f(x))$. For example, given a \code{sort} function, we might want to ensure that the output is always sorted and contains a permutation of its input elements. The key idea of property-based testing is to check such generic properties using auto-generated inputs values via random sampling of the input space $X$. 

A special class of property-based testing checks properties that should hold across multiple invocations of some program function when the inputs are changed in a specified manner---this is called \emph{metamorphic testing}~
\cite{Chen98}. For example, given a \code{sort} function, we can check whether \code{sort(x)} and \code{sort(reverse(x))} are equal for all values \code{x}.

Meta-reasoning tasks in statutory or contract law often deal with ensuring that the legal text satisfies certain consistency properties. For example, a desirable property of a tax legislation may be that the same income dollar is not taxed twice, or that the same expense dollar is not allowed as a deduction in two different ways. In estate planning, consider a complex \emph{last will and testament} having many conditional branches to decide how assets should be distributed depending on which beneficiaries are alive at the time of the will's execution; the testator might want to ensure that beneficiary $X$ always receives more than beneficiary $Y$ regardless of the circumstances and how the conditional branches are resolved. Such meta-reasoning is thus a form of edge-case analysis.

We argue that meta-reasoning tasks can potentially be automated using the key principles of metamorphic property-based testing and combining these with AI-driven techniques of interpretation. For example, consider \hyperref[irc:63c7B]{§63(c)(7)(B)} from the IRC (ref. Appendix~\ref{app:irc}), which provides adjustments for inflation to the basic standard deduction applicable in tax years 2018--2025. One may want to analyze the code to ask: \emph{Can the basic standard deduction available for single filers ever decrease year-over-year in this period?} Perhaps a legislator or other stakeholder is interested in understanding the effects of year-over-year deflation on the tax code. Answering this question involves not just interpreting the statutes but also reasoning about how they would apply in hypothetical edge-case scenarios.

Unfortunately, asking an LLM to analyze general properties of legal statutes directly does not always produce accurate results. When directly prompted with the aforementioned question, Claude~3 Opus determines ``\hyperref[irc:63c7Bii]{§63(c)(7)(B)(ii)} states that for taxable years beginning after 2018, these \$18,000 and \$12,000 amounts shall each be increased by a cost-of-living adjustment. [...] Since the amounts can only be increased and not decreased according to this special rule for 2018-2025, the basic standard deduction \emph{cannot decrease} year-over-year within this period.'' (emphasis added). However, it turns out that this is not fully correct.

We can help Claude by expressing our question as a metamorphic property-based test to verify the \emph{monotonicity} of the inflation adjustments---Given any two years $X$ and $Y$, such that $2018 \leq X < Y \leq 2025$, and hypothetical C-CPI-U values for the correspondingly preceding calendar years $I_x$ and $I_y$ in the numeric range [100, 200], calculate $D(X, I_x)$ and $D(Y, I_y)$ respectively, where function $D$ is the computation of basic standard deduction for single filers in a given year assuming a given C-CPI-U value for the preceding calendar year; then check whether $D(X, I_x) \leq D(Y, I_y)$ is true. Now do this for random values of $X, Y, I_x, I_y$ in a loop---in each iteration, use the LLM to calculate the inflation adjustment using the prompts shown earlier.

It takes about 2.4 iterations on average to discover that it is indeed possible to violate the stated property. Although the adjustments defined in \hyperref[irc:63c7B]{§63(c)(7)(B)} and \hyperref[irc:1f3]{§1(f)(3)} do indeed only allow for \emph{increases} to the basic standard deduction value listed in \hyperref[irc:63c2C]{§63(c)(2)(C)} (which is ``\$12,000'' in the period 2018--2025), the amount by which this value is increased can seemingly decrease year-over-year if there is deflation during those subsequent years. Property-based testing can reveal a situation where the monotonicity property is violated: for instance, $D(2024, 168.1)$=\$14,550 and $D(2025, 167.0)$=\$14,500; that is, if there was a hypothetical 0.65\% deflation in 2024, then the basic standard deduction for single filers in tax year 2025 as computed by \hyperref[irc:63c2C]{§63(c)(2)(C)} and \hyperref[irc:1f3]{§1(f)(3)} appears to be \$50 lower than that computed for the previous year.

Property-based testing thus allows us to reason about invariants across many hypothetical scenarios.

\section{Other Software Engineering Applications}

Given our view of legal reasoning through the lens of software engineering, one can imagine applications of a number of other software analysis techniques.

\paragraph{Example minmization} Delta debugging~\cite{Zeller02} is a technique for isolating the root cause of a program failure in terms of input data (e.g., think of a large file that causes a word processor to crash). The key idea is to repeatedly slice and dice the input until we obtain the smallest possible chunk that also reproduces the failure. 
    
In legal reasoning or meta-reasoning, delta debugging can be applied to any task where the output is an example scenario (ref. Sections~\ref{sec:example-generation} and \ref{sec:pbt}), and the objective is to find the \emph{smallest example such that a particular rule applies in a specific way}. We only need to implement support for input \emph{reduction} operators, which for natural language inputs (such as Examples~\ref{ex:interpretation}--\ref{ex:mutation-testing}) might mean deleting individual sentences or logical clauses separated by conjunctions. After that, input minimization via the $ddmin$ algorithm~\cite{Zeller02} is simply an application of interpretation (ref.~Section~\ref{sec:substitution-inlining}) in a loop, with the use of coverage analysis (ref.~Section~\ref{sec:coverage}) to determine whether reduced inputs exercise given rules (i.e., the \emph{oracle} of reduction).
    
\paragraph{Amendment validation} 
    Consider the meta-reasoning task of understanding the scope of \emph{amendments} to legislation or contracts, such as a bill introduced by the U.S. Congress that reforms the tax code, or a customized rider attached to an insurance policy. 
    
    In software, {change impact analysis}~\cite{Arnold96} is a technique for identifying which modules in a software get affected by a small patch applied to the program. The basic idea is to build a directed graph of various program modules and their inter-dependencies using both static source code information and dynamic hints from previous test executions. The impact of changes can then be identified by following paths through such a graph. 
    
    In legal reasoning, it is already possible to construct such static dependency graphs by mining explicit cross-references between various sections~\cite{DeMaat06, Sannier17}. With AI-driven interpretation and example generation with coverage analysis (ref.~Section~\ref{sec:coverage}), we can also add test cases as nodes in this graph. An amendment that modifies existing rules could then be analyzed by re-running ``test cases'' whose coverage directly or indirectly depends on the modified rule. Just like with the mutation testing approach~(ref.~Section~\ref{sec:mutation-testing}), the goal here would be to identify example scenarios (i.e., the ``test cases'') which are evaluated differently under the original and the amended rules.

\paragraph{Amendment composition} A stretch goal for using AI in legal analysis would be to prototype new amendments that have a specific impact on selected input scenarios.

In software engineering, the field of \emph{automatic program repair} (APR)~\cite{LeGoues21} develops techniques for fixing bugs by synthesizing appropriate patches. In a classical formation, APR starts with a program and a suite of tests such that one bug-revealing test fails. The task is to find a software patch such that this test and all the other test cases in the original suite pass with the new program. The key idea behind search-based APR techniques is to search over a space of small changes to the program until all the tests pass in the expected way.
    
Generative AI could potentially be used to brainstorm amendments to legal texts (the ``patches'') having a desired effect by providing natural language test scenarios such that certain rules should apply in desirable ways. We expect this application to be more useful in prototyping changes to contracts between business entities rather than drafting legislative bills.

\section{Discussion}

\subsection{Threats to Validity}

A key challenge with using LLMs to perform any sort of reasoning or analysis task is to ensure the reliability of their responses. LLMs are notorious for sometimes hallucinating~\cite{Maynez20-Hallucination} or providing incorrect responses for trivial reasoning tasks~\cite{Blair-Stanek23-GPT3Statutory}. While we do not offer any silver bullets in this paper, we note that this is a problem being actively researched by the AI community~\cite{Ji23-Survey, Tonmoy24}; mitigation strategies include extracting explanations for traceability~\cite{Dalvi21-EntailmentTrees} and refining answers via self-reflecting feedback loops~\cite{Ji23-Mitigation, Dhuliawala23}.

One common failure mode we have observed pertains to cases when the LLM is pre-trained on legal documents from the domain. For example, when working with U.S. tax law, we can be certain that commercial LLMs such as Claude would have been trained on the official IRC statutes as well as thousands of web articles discussing contemporary tax issues. Consequently, we have found that applying techniques such as mutation testing or presenting questions about hypothetical new laws sometimes results in sub-optimal outcomes because the model continues to ``believe'' the current law instead of new rules or situations that we may provide in prompts. In future work, we plan to investigate a mitigation technique where key domain-specific terms in the statutes (such as ``taxable income'') are replaced by domain-agnostic terms (such as ``foobar'') in order to disconnect the model's grounding in background knowledge.

\subsection{Alternative Approaches and Related Work}

In this paper, most of the examples considered directly prompting an LLM to perform numeric calculations based on natural language text in legal statutes. An alternative to this approach is to first prompt the LLM to synthesize an executable program that performs the same computation and then to execute such a program with inputs from a given scenario. We attempted to do so using Catala~\cite{Merigoux21-Catala}, a domain-specific programming language for formalizing statutes, as well as regular Python; however, our initial experience with this approach was quite unsatisfactory. One main problem we encountered is that legal statutes do not have a clear ``entry point'' and linear flow of logic; moreover, they do not upfront declare which terms may be re-interpreted by other sections. The task of manually transcribing statutes into Catala or Python programs often requires making multiple passes over the program to alter definitions. For example, when some rule $R$ requires applying section $S$ but replacing the term $X$ in place of $Y$, we need to go back to the definition of $S$ and convert constants or expressions into variables or subroutines, respectively, so that we may plug-in $X$ as an input. In our experience, we found current LLMs to not be well-suited for this successive non-linear refinement of programs with standard prompting techniques, though future research may discover more effective solutions.

Holzenberger and Van Durme~\cite{Holzenberger21-StatutoryReasoning, Holzenberger23-SymbolicStatutoryReasoning} have proposed performing statutory reasoning by extracting the structure of legal statutes in the form of graphs and Prolog programs---the approach appears to be effective, but the representations are not designed to be easy for humans to read and debug. The software engineering techniques presented in our paper, such as coverage analysis, mutation testing, or property-based testing, could potentially be applied on top of these Prolog programs as well.

Metamorphic testing has previously been applied to tax preparation software~\cite{TizpazNiari23-MetamorphicTaxPrep}; it would be interesting to validate the metamorphic properties used in this prior work by applying them directly on the underlying natural-language statutes. 

\subsection{Implications and Aspirations of the Work}

How will the techniques proposed in this paper apply in the real world? First, we want to clarify that we do not expect AI-driven legal analysis tools to completely replace any individual task performed by lawyers, primarily due to the lack of reliability in any statistical model. Instead, we anticipate that a software engineering approach to legal reasoning will enable the development of search \& analysis tools that augment current legal processes that are performed predominantly manually.

For lawmakers and other legal professionals involved in drafting laws, policies, or contracts, SE+AI-enabled tools could provide useful insights into the breadth and limitations of specific legal rules under consideration. Such tools can help drafters avoid producing inconsistent documents or making unintentional mistakes.

For ordinary citizens, legal analysis tools could potentially help provide useful insights that are only otherwise available to the more resourceful clients who can afford to retain legal experts. For example, we can imagine tools that can help individuals compare and demystify documents such as insurance policies, rental leases, and credit card agreements by providing ``what-if'' example generation and edge-case analysis capabilities; while not a substitute for legal advice, these insights could potentially form the basis for subsequent negotiation unlocked by the increased transparency. 

\section{Conclusion}

Generative AI has unlocked new ways of automating deductive legal reasoning. The intricacies and peculiarities of legalese make it much more suitable to treat statutes and contracts as computer programs rather than as natural language databases of information. Consequently, we argue that this problem can benefit from the principled application of techniques well-studied in the field of software engineering such as code coverage, function inlining, mutation testing, metamorphic property-based testing, and more. We hope that this paper forms the basis for researchers to investigate further synergies across software analysis and computational law.

\begin{acks}
The author would like to thank the anonymous reviewers of the \emph{Onward!} track for their valuable comments and suggestions for improvement. The paper also benefited from feedback provided by Sarah Lawsky, Travis Breaux, Sarah Santos, Vasudev Vikram, and Sam Estep. Finally, the author thanks Bogdan Vasilescu for encouraging him to pursue and publish these ideas.
\end{acks}

\appendix

\section{Excerpt from Internal Revenue Code}
\label{app:irc}

In this appendix, we list excerpts from 26 U.S.C. \hyperref[irc:63]{§63}, \hyperref[irc:1]{§1}, and \hyperref[irc:151]{§151} (as of April 2024) that are provided to an LLM for reasoning about tasks described in the sections above.

\begin{lstlisting}[style=statute]
@\textbf{§63. Taxable income defined} \label{irc:63}@

...

(b) Individuals who do not itemize their deductions @\label{irc:63b}@

  In the case of an individual who does not elect to itemize his deductions for the taxable year, for purposes of this subtitle, the term “taxable income” means adjusted gross income, minus—
  
  (1) the standard deduction,@\label{irc:63b1}@

  ...

(c) Standard deduction @\label{irc:63c}@

    (1) In general @\label{irc:63c1}@

    Except as otherwise provided in this subsection, the term "standard deduction" means the sum of-

        (A) the basic standard deduction, and @\label{irc:63c1A}@

        (B) the additional standard deduction. @\label{irc:63c1B}@

    (2) Basic standard deduction @\label{irc:63c2}@

    For purposes of paragraph (1), the basic standard deduction is-

        (A) 200 percent of the dollar amount in effect under subparagraph (C) for the taxable year in the case of-@\label{irc:63c2A}@

            (i) a joint return, or @\label{irc:63c2Ai}@

            (ii) a surviving spouse (as defined in section 2(a)), @\label{irc:63c2Aii}@

        (B) $4,400 in the case of a head of household (as defined in section 2(b)), or @\label{irc:63c2B}@

        (C) $3,000 in any other case. @\label{irc:63c2C}@

    (3) Additional standard deduction for aged and blind @\label{irc:63c3}@

    For purposes of paragraph (1), the additional standard deduction is the sum of each additional amount to which the taxpayer is entitled under subsection (f).

    (4) Adjustments for inflation @\label{irc:63c4}@

        In the case of any taxable year beginning in a calendar year after 1988, each dollar amount contained in paragraph (2)(B), (2)(C), or (5) or subsection (f) shall be increased by an amount equal to— 
        
        ...

    (7) Special rules for taxable years 2018 through 2025 @\label{irc:63c7}@

    In the case of a taxable year beginning after December 31, 2017, and before January 1, 2026-

    (A) Increase in standard deduction @\label{irc:63cA}@

        Paragraph (2) shall be applied—

        (i) by substituting “$18,000” for “$4,400” in subparagraph (B), and @\label{irc:63c7Ai}@

        (ii) by substituting “$12,000” for “$3,000” in subparagraph (C). @\label{irc:63c7Aii}@
        
    (B) Adjustment for inflation @\label{irc:63c7B}@
            
        (i) In general @\label{irc:63c7Bi}@
    
        Paragraph (4) shall not apply to the dollar amounts contained in paragraphs (2)(B) and (2)(C).

        (ii) Adjustment of increased amounts @\label{irc:63c7Bii}@
        
            In the case of a taxable year beginning after 2018, the $18,000 and $12,000 amounts in subparagraph (A) shall each be increased by an amount equal to—
            
            (I) such dollar amount, multiplied by
            
            (II) the cost-of-living adjustment determined under section 1(f)(3) for the calendar year in which the taxable year begins, determined by substituting “2017” for “2016” in subparagraph (A)(ii) thereof.

            If any increase under this clause is not a multiple of $50, such increase shall be rounded to the next lowest multiple of $50.

...

(f) Aged or blind additional amounts @\label{irc:63f}@

  (1) Additional amounts for the aged @\label{irc:63f1}@
  
    The taxpayer shall be entitled to an additional amount of $600—
    
    (A) for himself if he has attained age 65 before the close of his taxable year, and @\label{irc:63f1A}@
    
    (B) for the spouse of the taxpayer if the spouse has attained age 65 before the close of the taxable year and an additional exemption is allowable to the taxpayer for such spouse under section 151(b). @\label{irc:63f1B}@

  ...

@\textbf{§1. Tax imposed}  \label{irc:1}@

...

(f) Adjustments in tax tables so that inflation will not result in tax increases @\label{irc:1f}@

    ...

    (3) Cost-of-living adjustment @\label{irc:1f3}@

        For purposes of this subsection—

        (A) In general @\label{irc:1f3A}@

            The cost-of-living adjustment for any calendar year is the percentage (if any) by which—

            (i) the C-CPI-U for the preceding calendar year, exceeds @\label{irc:1f3Ai}@

            (ii) the CPI for calendar year 2016, multiplied by the amount determined under subparagraph (B). @\label{irc:1f3Aii}@

        (B) Amount determined @\label{irc:1f3B}@

            The amount determined under this clause is the amount obtained by dividing—

            (i) the C-CPI-U for calendar year 2016, by @\label{irc:1f3Bi}@

            (ii) the CPI for calendar year 2016. @\label{irc:1f3Bii}@

        (C) Special rule for adjustments with a base year after 2016 @\label{irc:1f3C}@
            
        For purposes of any provision of this title which provides for the substitution of a year after 2016 for “2016” in subparagraph (A)(ii), subparagraph (A) shall be applied by substituting “the C-CPI-U for calendar year 2016” for “the CPI for calendar year 2016” and all that follows in clause (ii) thereof.

    (4) CPI for any calendar year @\label{irc:1f4}@
        
    For purposes of paragraph (3), the CPI for any calendar year is the average of the Consumer Price Index as of the close of the 12-month period ending on August 31 of such calendar year.

    (6) C-CPI-U @\label{irc:1f6}@
    
        ...

        (B) Determination for calendar year  @\label{irc:1f6B}@
    
        The C-CPI-U for any calendar year is the average of the C-CPI-U as of the close of the 12-month period ending on August 31 of such calendar year.

        ...

@\textbf{§151. Allowance of deductions for personal exemptions}  \label{irc:151}@

    (a) Allowance of deductions
    
    In the case of an individual, the exemptions provided by this section shall be allowed as deductions in computing taxable income.

    (b) Taxpayer and spouse @\label{irc:151b}@
    
    An exemption of the exemption amount for the taxpayer; and an additional exemption of the exemption amount for the spouse of the taxpayer if a joint return is not made by the taxpayer and his spouse, and if the spouse, for the calendar year in which the taxable year of the taxpayer begins, has no gross income and is not the dependent of another taxpayer.

    ...
\end{lstlisting}

\section{Inflation Data for Cost-of-Living Adjustments}
\label{app:cpi}
\label{app:inflation}

The following table provides (hypothetical) values for the Chained Consumer Price Index for All Urban Consumers (C-CPI-U) for calendar years 2017--2024, calculated using the averaging methodology in \hyperref[irc:1f6B]{\S 1(f)(6)(B)}.

\begin{table}[!h]
\footnotesize
\begin{tabular}{r|r|r|r|r|r|r|r|r}
    \toprule
    Year & 2017 & 2018 & 2019 & 2020 & 2021 & 2022 & 2023 & 2024 \\
    \midrule
    Value & 138.2 & 141.0 & 143.2 & 144.8 & 149.1 & 159.9 & 168.3 & 172.5 \\
    \bottomrule
\end{tabular}
\end{table}

\section{Sample Transcripts}

This section contains some sample responses from Anthropic's Claude 3 Opus when provided examples from the paper. The detailed transcripts with full prompts are available at \url{https://github.com/cmu-pasta/onward24}.

\subsection{Determining Standard Deduction without Inflation Adjustments}

When given the system prompt (ref. Fig.~\ref{fig:system-prompt}) followed by a prompt containing Example~\ref{ex:interpretation} and the phrase ``Let's think step by step'' for zero-shot chain-of-thought~\cite{Kojima22-ZeroShotCoT}, Claude responds:

\begin{lstlisting}[style=llm]
Okay, let's break this down step-by-step based on the given statutes:

1. Determine the standard deduction for Alice and Bob in 2018:
   - According to §63(c)(7)(A), for taxable years 2018 through 2025, the basic standard deduction for a joint return is $24,000 ($18,000 substituted for $4,400 in §63(c)(2)(B), and $12,000 substituted for $3,000 in §63(c)(2)(C)).

2. Determine if Alice and Bob are entitled to any additional standard deduction amounts:
   - Per §63(f)(1)(A) and (B), taxpayers are entitled to an additional $600 each if they have attained age 65 before the close of the taxable year.
   - In 2018, Alice was 37 years old (born in 1981), and Bob was 43 years old (born in 1975). Neither had attained age 65.
   - Therefore, they are not entitled to any additional standard deduction amounts.

3. Calculate their total standard deduction:
   - Their total standard deduction is the basic standard deduction of $24,000, as they do not qualify for any additional standard deduction amounts.

4. Calculate their taxable income:
   - According to §63(b), for individuals who do not itemize their deductions, taxable income means adjusted gross income minus the standard deduction.
   - Alice and Bob's adjusted gross income in 2018 was $216,350.
   - Their taxable income = $216,350 - $24,000 = $192,350.

Therefore, Alice and Bob's taxable income for 2018 is $192,350.
\end{lstlisting}

\subsection{Calculating Inflation Adjustments to Standard Deduction without Inlining}
\label{app:claude-inflation_baseline}

When asked to solve Example~\ref{ex:substitution} with only the system prompt (ref. Fig.~\ref{fig:system-prompt}), inflation data (ref. Appendix~\ref{app:cpi}), and using zero-shot chain-of-thought~\cite{Kojima22-ZeroShotCoT}, Claude responds:

\begin{lstlisting}[style=llm]
Okay, let's break this down step-by-step based on the information provided in the statutes:

1. Determine the basic standard deduction:
   - According to §63(c)(7)(A), for taxable years 2018 through 2025, the basic standard deduction for a joint return is $18,000.

2. Adjust the basic standard deduction for inflation:
   - §63(c)(7)(B)(ii) states that for taxable years beginning after 2018, the $18,000 amount shall be increased by the cost-of-living adjustment determined under §1(f)(3), substituting "2017" for "2016" in §1(f)(3)(A)(ii).
   - The cost-of-living adjustment for 2025 is calculated as follows:
     * C-CPI-U for 2024 (preceding year): 172.5
     * C-CPI-U for 2017: 138.2
     * Adjustment = (172.5 / 138.2) - 1 = 0.2482 (24.82%)
   - Increase the $18,000 basic standard deduction by 24.82%:
     * $18,000 x (1 + 0.2482) = $22,467.60
   - Round down to the nearest multiple of $50:
     * Adjusted basic standard deduction = $22,450

3. Determine the additional standard deduction:
   - According to §63(f)(1)(A) and (B), taxpayers are entitled to an additional $600 each if they have attained age 65 before the close of the taxable year.
   - In 2025, Alice is 44 years old, and Bob is 49 years old, so they are not entitled to the additional standard deduction.

Therefore, for the tax year 2025, Alice and Bob's standard deduction is $22,450.
\end{lstlisting}

This answer is incorrect. Although Claude correctly calculates the inflation adjustment of 24.82\%, it multiplies this with a figure of \$18,000 (corresponding to the basic standard deduction for a head-of-household) instead of multiplying by \$12,000 and then doubling it for a joint return.

\subsection{Coverage Tracing}
\label{app:claude-coverage}

The following is a user prompt for coverage tracing on Example~\ref{ex:interpretation}---the usual system prompt containing the statutes (ref. Fig.~\ref{fig:system-prompt}) was also provided.

\begin{lstlisting}[style=llm]
Alice and Bob are married and filing taxes jointly for the year 2018. Alice was born on 1/1/1981 and Bob was born on 12/30/1975. In 2018, Alice and Bob's adjusted gross income was $216,350. Alice and Bob do not itemize their deductions and do not qualify for any deductions other than the standard deduction. What is their taxable income for 2018?

Let's think step by step. In each step, reference the section or paragraph number and justify your reasoning. If you encounter any cross-references to other sections or paragraphs, perform the reasoning in a sub-list and return to the previous step. Do this recursively for any subsequent cross-references. If you encounter any substitutions of values, perform the substitution and reproduce the substituted sentences verbatim before deciding if they actually apply. End your response with two special lines as follows:

Answer: <dollar amount calculated>
Coverage: <comma-separated list of rules referenced, identified by their most specific section/paragraph numbers>
Relevance: <comma-separated list of rules that applied to this case because the conditions were true, again identified by their most specific section/paragraph numbers>
\end{lstlisting}

\noindent Claude's response:

\begin{lstlisting}[style=llm]
Okay, let's break this down step-by-step:

1. According to §63(b), since Alice and Bob do not itemize their deductions, their taxable income is their adjusted gross income minus the standard deduction. 
   - They have an adjusted gross income of $216,350.
   - We need to calculate their standard deduction.

2. §63(c)(1) says the standard deduction is the sum of the basic standard deduction and the additional standard deduction.

3. For the basic standard deduction, §63(c)(2) provides dollar amounts, but §63(c)(7) says for a taxable year beginning after December 31, 2017 and before January 1, 2026, paragraph (2) shall be applied with some substitutions.
   1. §63(c)(7)(A)(i) says to substitute "$18,000" for "$4,400" in §63(c)(2)(B).
      - After substitution, §63(c)(2)(B) reads: "$18,000 in the case of a head of household (as defined in section 2(b))". This does not apply as Alice and Bob are filing jointly, not as head of household.
   2. §63(c)(7)(A)(ii) says to substitute "$12,000" for "$3,000" in §63(c)(2)(C).
      - After substitution, §63(c)(2)(C) reads: "$12,000 in any other case". 
   3. §63(c)(2)(A) says the basic standard deduction is 200% of the amount in §63(c)(2)(C) for a joint return. 
      - Since §63(c)(2)(C) is $12,000 after the substitution, the basic standard deduction for Alice and Bob is $24,000.

4. For the additional standard deduction, §63(c)(3) says it is the sum of the additional amounts the taxpayer is entitled to under §63(f).
   1. §63(f)(1)(A) provides an additional amount of $600 if the taxpayer has attained age 65 before the close of the taxable year. Bob was born on 12/30/1975, so he was 43 in 2018 and this does not apply. Alice was born on 1/1/1981, so she was 37 in 2018 and this also does not apply.
   2. §63(f)(1)(B) provides an additional $600 for the spouse of the taxpayer if the spouse has attained age 65 and an additional exemption is allowable under §151(b). Neither condition is met (Bob is under 65, and they are filing jointly so §151(b) does not provide an additional exemption). So this also does not apply.
   - Therefore, Alice and Bob are not entitled to any additional standard deduction under §63(f).

5. The total standard deduction is thus just the basic standard deduction of $24,000.

6. Going back to §63(b), the taxable income is the adjusted gross income of $216,350 minus the standard deduction of $24,000.

Therefore:

Answer: $192,350
Coverage: §63(b), §63(c)(1), §63(c)(2), §63(c)(7), §63(c)(7)(A)(i), §63(c)(7)(A)(ii), §63(c)(2)(A), §63(c)(2)(C), §63(c)(3), §63(f), §63(f)(1)(A), §63(f)(1)(B), §151(b) 
Relevance: §63(b), §63(c)(1), §63(c)(7), §63(c)(7)(A)(ii), §63(c)(2)(A), §63(c)(2)(C)
\end{lstlisting}

\end{document}